\documentclass[reprint,aip,amsmath,amssymb,nofootinbib]{revtex4-2}
\usepackage{amssymb}
\usepackage{amssymb}
\usepackage{amssymb}
\usepackage{graphicx}
\usepackage{graphics}
\usepackage{amsmath}
\usepackage{placeins}
\usepackage{hyperref}
\usepackage[dvipsnames]{xcolor}
\usepackage{wasysym}
\usepackage{soul}
\usepackage{float}
\usepackage{array}
\usepackage{tabulary}
\usepackage{caption}
\usepackage{fancyhdr}
\usepackage{wrapfig}
\usepackage{multirow}

\usepackage[normalem]{ulem}
\usepackage{graphicx}
\usepackage{color,soul}
\usepackage{subfigure}
\usepackage{dcolumn}
\usepackage{bm}
\usepackage{hyperref}
\usepackage{url}
\usepackage{multirow}
\hypersetup{
    colorlinks=true,
    linkcolor=purple,
    filecolor=blue,      
    urlcolor=blue,
    citecolor=blue,
}
\usepackage{float}

\begin{document}
	
	\title{FEbeam: Cavity and Electron Emission Data Conversion, Processing and Analysis. A Freeware Toolkit for RF Injectors}
	
	\author{\firstname{Mitchell} \surname{Schneider}}
	\email{schne525@msu.edu}
    \affiliation{Department of Electrical and Computer Engineering, Michigan State University, MI 48824, USA}
	\affiliation{Department of Physics and Astronomy, Michigan State University, MI 48824, USA}
   	\author{\firstname{Emily} \surname{Jevarjian}}
	\email{jevarji1@msu.edu}
	\affiliation{Department of Physics and Astronomy, Michigan State University, MI 48824, USA}
    \affiliation{Department of Electrical and Computer Engineering, Michigan State University, MI 48824, USA}
    \author{\firstname{Jiahang} \surname{Shao}}
    \email{jshao@anl.gov}
	\affiliation{High Energy Physics Division, Argonne National Laboratory, IL 60439, USA}
	\author{\firstname{Sergey V.} \surname{Baryshev}}
	\email{serbar@msu.edu}
	\affiliation{Department of Electrical and Computer Engineering, Michigan State University, MI 48824, USA}
	
	
	\begin{abstract}
	FEbeam is a compact field emission data processing interface with the capability to analyze the field emission cathode performance in an rf injector by extracting the field enhancement factor, local field, and effective emission area from the Fowler-Nordheim equations. It also has the capability of processing beam imaging micrographs using its sister software, FEpic. The current version of FEbeam was designed for the Argonne Cathode Teststand (ACT) of the Argonne Wakefield Accelerator facility switch yard. With slight modifications, FEbeam can work for any rf field emission injector. This software is open-source and can be found at \href{https://github.com/schne525/FEbeam}{GitHub}.
	\end{abstract}
	
	
	\maketitle
	
	\section*{Introduction}
	
\begin{figure*}[htp]
		\includegraphics[width=17cm]{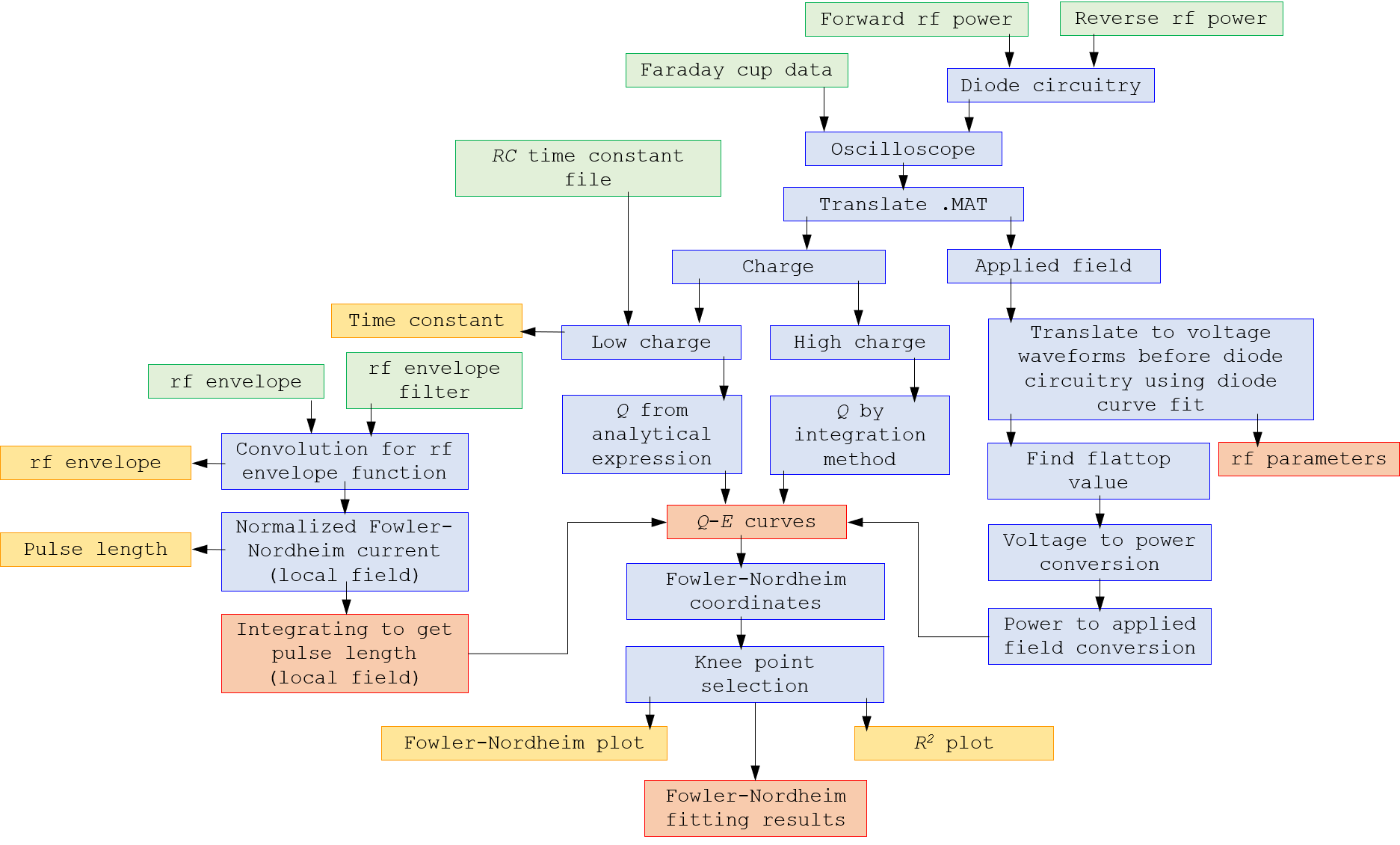}
		\caption{\label{1} FEbeam’s workflow map showing how the data processing algorithm (in blue) takes the raw data from input files (in green) to the output data (in red) and the output figures (in yellow).}
	\end{figure*}

As field emitters are poised to be the next generation of electron sources in rf and microwave systems, the ability to analyze field emission cathode performance by processing extensive raw data in a compact, easy-to-use format is crucial. Therefore, a toolbox called FEbeam was developed to fill the existing gap. FEbeam consists of 17 different Matlab functions. A single graphical user interface (GUI) conveniently combines and cross-links them together. Fig.~\ref{1} shows the workflow map of all of the components necessary to take the raw data and convert it to analyze and compare the field emission cathode performance. The green boxes in Fig.~\ref{1} are the input data and parameter files, the blue boxes represent a simplified version of the internal pipeline for data processing, the red boxes are the output files, and the yellow boxes are the output figures.

This paper is laid out as follows: Section \ref{QE} describes how charge ($Q$)-electric field ($E$) curves are obtained; Section \ref{FN} describes how Fowler-Nordheim (FN) fitting parameters get determined; Section \ref{all} describes image and resulting scientific data postprocessing and plotting; Section \ref{conc} contains conclusion and outlook.
Given the direct link with the physics of dark current and vacuum breakdown/arc, additional sister toolkits FEbreak (breakdown statistics) and FEpic (image/pattern recognition) \cite{FEpic} are being finalized to be jointly used with FEbeam. Additional toolkit FEgen \cite{Fegen} exists for computational support.

	\section{\label{QE}Obtaining Q-E Curves}

\begin{figure}[htp]
		\includegraphics[width=8cm]{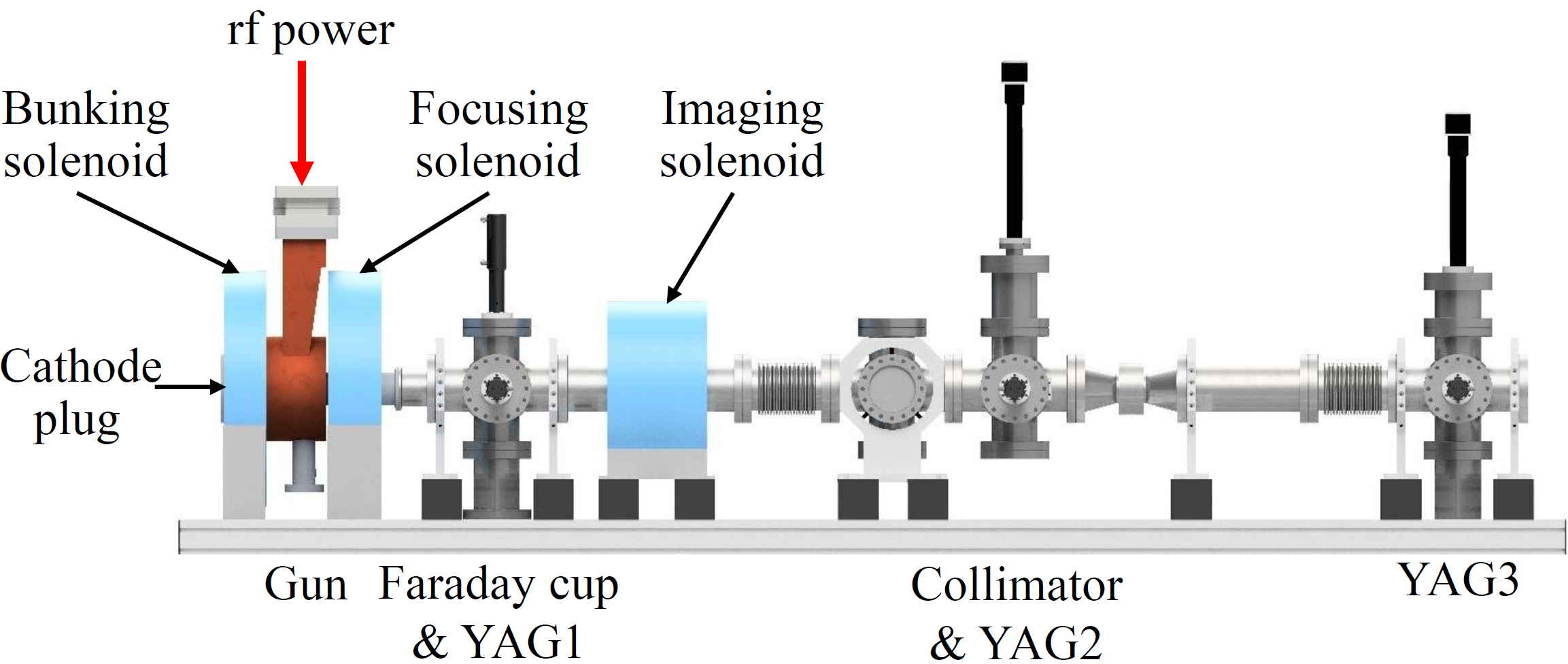}
		\caption{\label{2} Schematic of ACT beamline with rf pickups not shown, collimator denotes the 1 mm aperture.}
	\end{figure}

This FEbeam version is specifically optimized to automate the Argonne Cathode Teststand (ACT) shown in Fig.~\ref{2}, which a part of the Argonne Wakefield Accelerator (AWA) facility. The ACT is a 1/2 cell L-band (1.3 GHz) gun that can operate with field and photoemission sources with the maximum macroscopic gradient of 105 MV/m (that is for a planar cathode). The ACT has three yttrium aluminum garnet (YAG) screens which are used for imaging of the transverse electron distribution. At the location of YAG1, there is a dropdown Faraday cup to measure the charge out of the gun. At the location of YAG2, there is an exchangeable 1 mm aperture that is used to collimate the beam. Finally, fixed YAG3 is used for downstream electron beam imaging.

The forward and reverse power are measured using a directional coupler on the waveguide near klystron. The raw data used to calculate the field emission characteristics is obtained from the oscilloscope in a CSV format consisting of the voltage waveforms for the Faraday cup, forward power, and reverse power. Each CSV file is a single point in the $Q$-$E$ curve where each point consists of 10 individual pulse shots.

\begin{figure}[htp]
		\includegraphics[width=9cm]{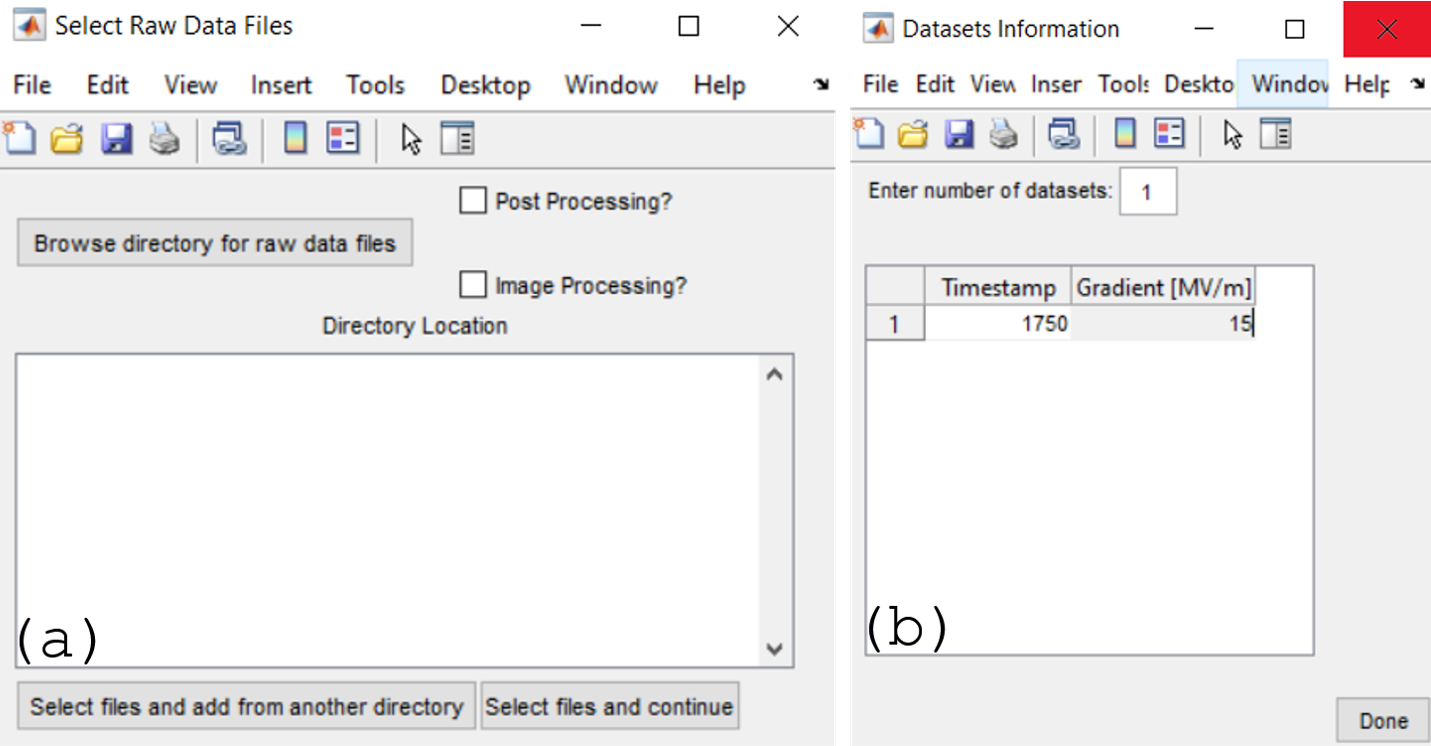}
		\caption{\label{3} (a) The initial window to select data to be processed; (b) Display screen that allows for the timestamp of the dataset to be associated with the conditioning gradient that will be later used to rename and group the files.}
	\end{figure}

The opening GUI screen, Select Raw Data Files, is shown in Fig.~\ref{3}a. This allows the user to select the folders in which the raw CSV files are located. The user then selects the files that they want to process in the GUI window. This screen has the options for image processing and postprocessing that are discussed in further detail in Section \ref{FN}. The default name setting for the CSV files groups is derived from the experimentally achieved conditioning gradient. For ease-of-use, FEbeam renames the files based upon the conditioning gradient for that dataset as shown in Fig.~\ref{3}b, Datasets Information. The conditioning field is the desired field that the cathode was conditioned to for a given dataset. Note that the conditioning field is used solely for naming convenience and that the applied field is the actual measured value.

\begin{figure}[htp]
		\includegraphics[width=7cm]{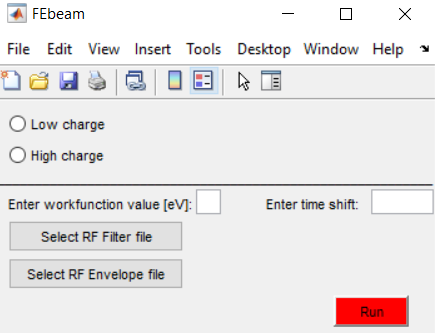}
		\caption{\label{4} The main interface of FEbeam. The top portion sets the rf parameters to translate the raw data into $Q$-$E$ curves. The lower portion calculates the pulse length of the function of the local field and calculates the emission envelope using the Fowler-Nordheim equations, based on the work function entered.}
	\end{figure}

After the datasets have been selected, renamed, and translated into Matlab data files (.MAT format), the rf parameters need to be set in the FEbeam screen, as shown in Fig.~\ref{4}. These parameters are used to translate the raw data into $Q$-$E$ curves after each individual file is grouped based on the conditioning gradient entered in Dataset Information (see Fig.~\ref{3}b). Parameters include options for a low charge or a high charge case scenario, and for selection of the rf filter and rf envelope needed to calculate the pulse length.

\begin{figure}[htp]
		\includegraphics[width=9cm]{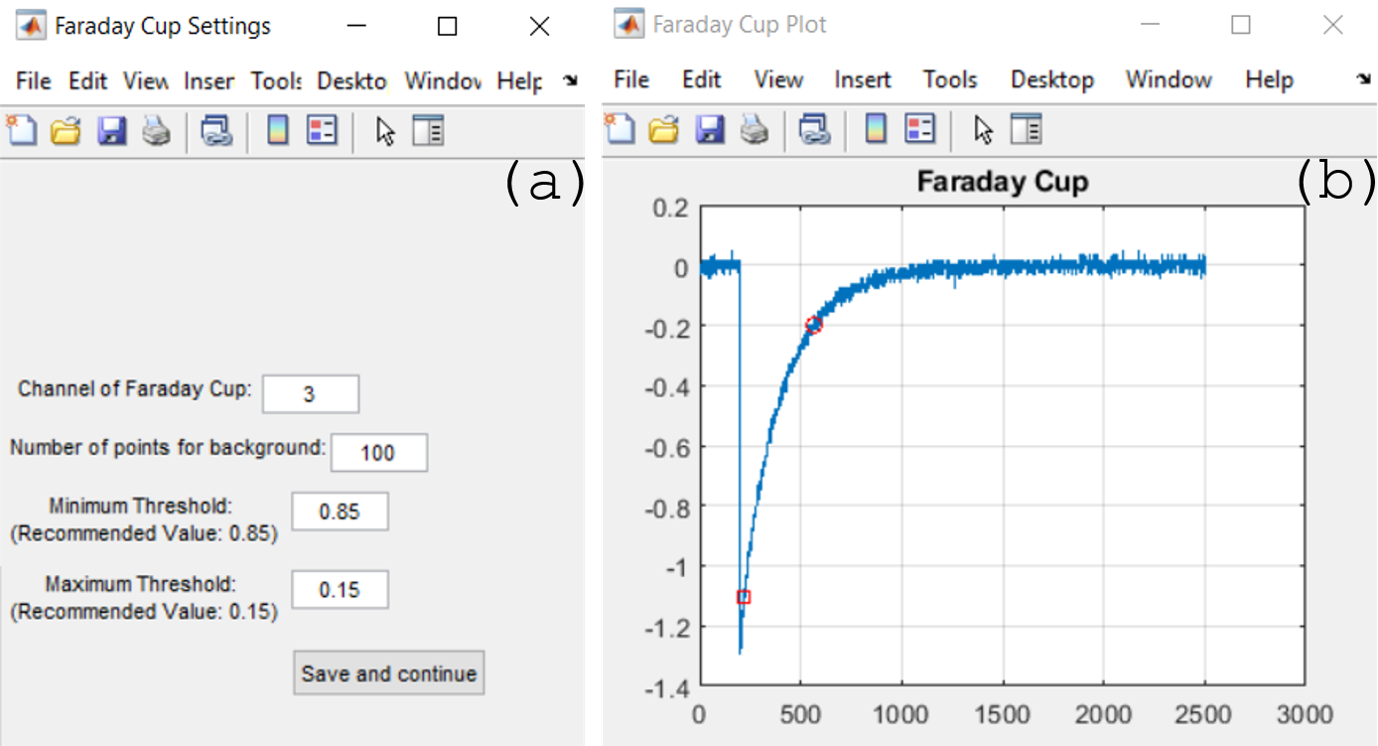}
		\caption{\label{5} (a) Faraday Cup Settings display interface for selecting the time constant for the low charge case. (b) Plot where red square and circle denote the location of the minimum and maximum threshold values corresponding to 90\% and 10\% of the maximum value.}
	\end{figure}

The difference between the high and low charge scenarios refers to the terminating impedance on the oscilloscope in the Faraday cup circuitry. The low charge case uses the impedance of 1 M$\Omega$ as the larger impedance allows for larger voltage drop read by the scope and hence higher sensitivity to lower charge beams. However, the Faraday cup signal length will increase by a few orders of magnitude. As a result rf pulse and Faraday cup signals cannot be captured simultaneously. Therefore, Faraday cup signal is recorded to only calculate $RC$ time constant $\tau_{RC}$ as illustrated by the interface screen shown in Fig.~\ref{5}. The charge in the Faraday cup for the low charge scenario can be then found using the top expression in Eq.~\ref{eq1}. When calculating the time constant, the user selects a range to use as back subtraction to set the noise floor and a range to calculate the $RC$ time constant. Recommended values of 90\% and 10\% are the minimum recommended to mitigate calculation errors and are illustrated in Fig.~\ref{5}.

\begin{figure}[htp]
		\includegraphics[width=9cm]{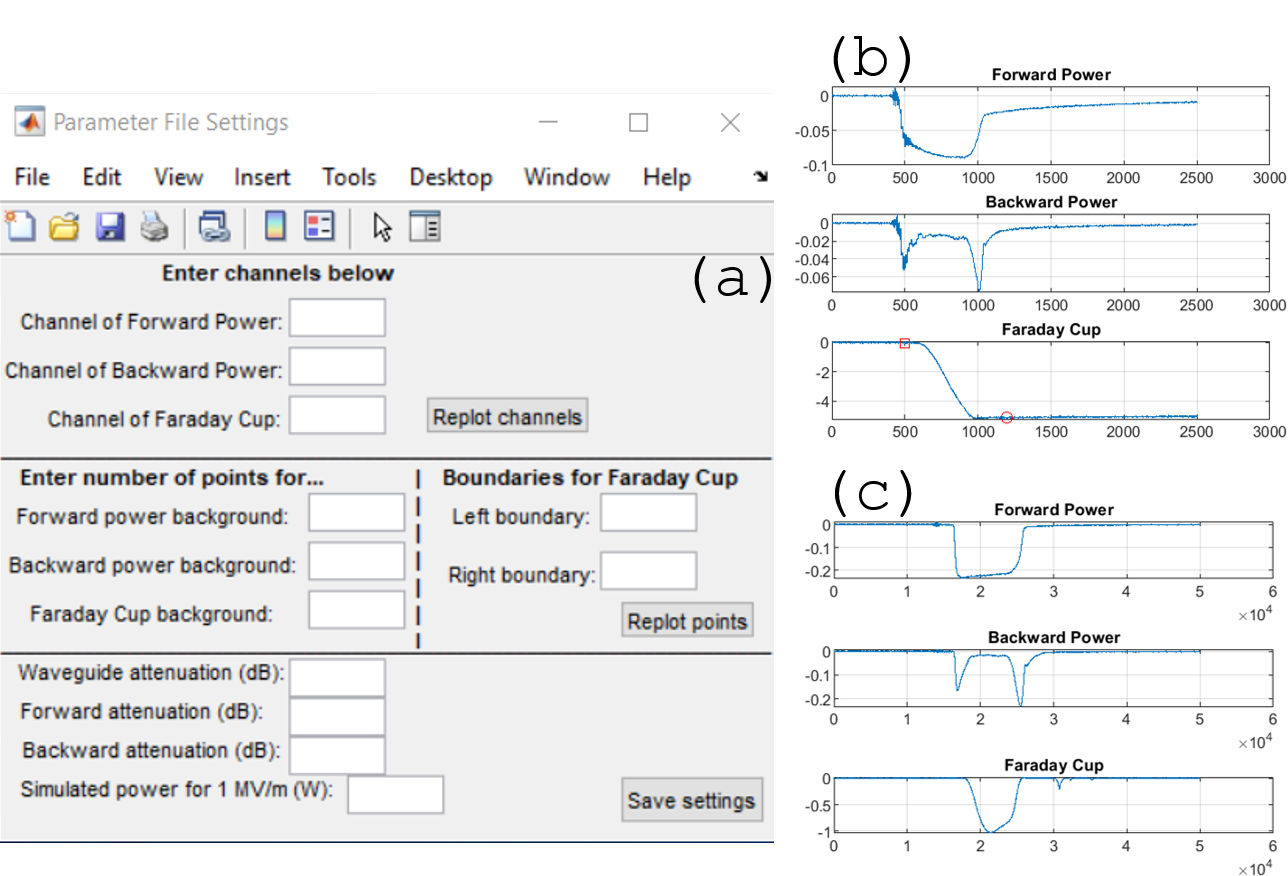}
		\caption{\label{6} (a) Parameter File Settings for entering the rf parameters to calculate the $Q$-$E$ curves; (b) waveforms for a low charge scenario; (c) waveforms for a high charge scenario.  Red marks denote the region that was considered in the Faraday cup to calculate the charge.}
	\end{figure}

In the high charge case, which uses the terminating impedance of 50 $\Omega$, the charge is directly calculated by integrating the signal from the Faraday cup over the range specified in the settings window (see Eq.~\ref{eq1}). In the Parameter File Settings (see Fig.~\ref{6}a), the operator selects which of the three channels represents the forward power, reverse power, and the Faraday cup signal. One gets to enter then which time integration range will be used to calculate the noise floor for the data. This integration range is used to calculate the charge collected. The final set of parameters associated with the specific rf system includes the attenuation factors due to the waveguides, the attenuators for the forward and reverse power, and the power to electric field conversion factor for the output power of a klystron.

In the ACT case, the waveguide attenuation is 0.2 dB, and the power conversion factor is 208.5 W of the input power corresponds to 1 MV/m of applied macroscopic field. The 208.5 W power conversion factor was obtained from SUPERFISH cavity simulation and is valid for planar cathode geometry only. The forward and reverse power attenuators, depending on the experiment set up, are either 10 or 15 dB with an addition of 0.5 dB added to account for the filter on the attenuator. The settings window is the same for the high and low charge cases: Fig.~\ref{6}b and Fig.~\ref{6}c show the Faraday cup waveforms for the low and high charge case relaying the basic difference. Summarizing, the charge is calculated as follows

\begin{equation}
 \begin{cases}
 
Q=-\frac{V^{mean}_{FC}}{Z_L}\tau_{RC}, & \text{low charge} \\

\\

Q=-\frac{1}{Z_L}\int_{t_1}^{t_2} V_{FC} dt, & \text{high charge}

\end{cases}
\label{eq1}
\end{equation}
where $V_{FC}$ is the voltage waveform from the Faraday cup within the specified boundaries in the settings, $\tau_{RC}$ is the time constant calculated for the low charge case, $Z_L$ is the load impedance, and $t_1$ and $t_2$ are the 10\% and 90\% boundaries.

If an oscilloscope with limited bandwidth is only available, it is still possible to measure envelope for the forward and reflected power in order to establish the applied field on the cathode. Fig.~\ref{6}b and Fig.~\ref{6}c demonstrate the result of this approach. For a narrow bandwidth scope, the raw forward and reverse rf signal picked up on the directional coupler installed on the waveguide near the L-band driving klystron was passed through a diode (Keysight 423B) circuit to modulate the frequency low enough that it can be read in by the oscilloscope. This step is not necessary for systems and facilities with access to a GHz scope, and can be bypassed if necessary. During data processing, the raw forward and reverse power waveforms from the oscilloscope are translated back into their original form by fitting the diode waveform to a 7-order polynomial. These conversion equations [Eq.~\ref{FP} and Eq.~\ref{RP}] for the ACT are as follows

\begin{equation}\label{FP}
\begin{aligned}    
\noindent P_{FP}^{original}=-0.034 + 0.0477\cdot V_{FP}+\\
+ 6.617\cdot 10^{-4}\cdot V_{FP}^2+\\
+ 4.948\cdot 10^{-6}\cdot V_{FP}^3-\\
- 5.411\cdot 10^{-8}\cdot V_{FP}^4+\\
+ 3.129\cdot 10^{-10}\cdot V_{FP}^5-\\
- 8.789\cdot 10^{-13}\cdot V_{FP}^6+\\
+ 9.657\cdot 10^{-16}\cdot V_{FP}^7   
\end{aligned}
\end{equation}

\begin{equation}\label{RP}
\begin{aligned}    
\noindent P_{RP}^{original}=-0.159 + 0.07052\cdot V_{RP}+\\
+ 2.4\cdot 10^{-5}\cdot V_{RP}^2+\\
+ 1.86\cdot 10^{-5}\cdot V_{RP}^3-\\
- 1.925\cdot 10^{-7}\cdot V_{RP}^4+\\
+ 1.079\cdot 10^{-9}\cdot V_{RP}^5-\\
- 3.028\cdot 10^{-12}\cdot V_{RP}^6+\\
+ 3.37\cdot 10^{-15}\cdot V_{RP}^7   
\end{aligned}
\end{equation}
where power waveforms for both the forward (FP) and reverse (RP) power have units of milliwatts in this equation.

To convert original voltage waveforms for the forward and reverse power into actual power, the following equations are used

\begin{equation}\label{forpower}
\begin{aligned} 
P_F = P_{FP}^{original}\cdot 10^{\big(\frac{ATT_{FP}-ATT_{WG}+ATT_{DC}}{10}-3\big)}
\end{aligned}
\end{equation}

\begin{equation}\label{revpower}
\begin{aligned} 
P_R = P_{RP}^{original}\cdot 10^{\big(\frac{ATT_{RP}+ATT_{WG}+ATT_{DC}}{10}-3\big)}
\end{aligned}
\end{equation}
where $ATT_{FP}$ is the forward power attenuator, $ATT_{RP}$ is the reverse power attenuator, $ATT_{WG}$ is the waveguide attenuation, and $ATT_{DC}$ is the attenuation for the directional coupler which, in this case, is 60 dB. The $-3$ term denotes the conversion from milliwatts to watts.

It is assumed that the rf pulse is a flat top; therefore, if we integrate over the forward power and then divide by half, this midpoint should be where the level of the flat top is, as shown by Eq.~\ref{flattop}. This corresponds to the median applied field over that rf pulse. Time $k$ corresponding to that midpoint is found from relation 

\begin{equation}\label{flattop}
\begin{aligned} 
\int_{t_i}^{t_k} P_F dt=\frac{1}{2}\int_{t_i}^{t_f} P_F dt
\end{aligned}
\end{equation}
where $t_i$ and $t_f$ are initial/start and final/end time integration boundaries.

Finally the applied macroscopic electric field can be found as
\begin{equation}\label{E-field}
\begin{aligned} 
E_c = \sqrt{\frac{P_F^k}{P_{factor}^{ACT}}}
\end{aligned}
\end{equation}
where $P_F^k$ is the power at time $t_k$ measured in W as found from Eq.~\ref{flattop} and $P_{factor}^{ACT}$ is the ACT power conversion factor equal to 208.5 W$\cdot$m$^2$/MV$^2$. The applied field and power conversion factor are imported from the Parameter File Settings, and labeled in accordance with the highest $E_c$ achieved, called $E_h$. Subsequently, each of the data points in the $Q$-$E$ curve are combined into a single file. The next step is converting the $Q$-$E$ curve into the Fowler-Nordheim coordinate representation to further extract the field emission characteristic parameters, the field enhancement factor $\beta$, cathode local or microscopic electric field $E_L = \beta \times E_c$, and the effective emission area $A_e$.

\section{\label{FN} Determining Fowler-Nordheim Parameters}
\begin{figure}[h!tbp]
		\includegraphics[width=8cm]{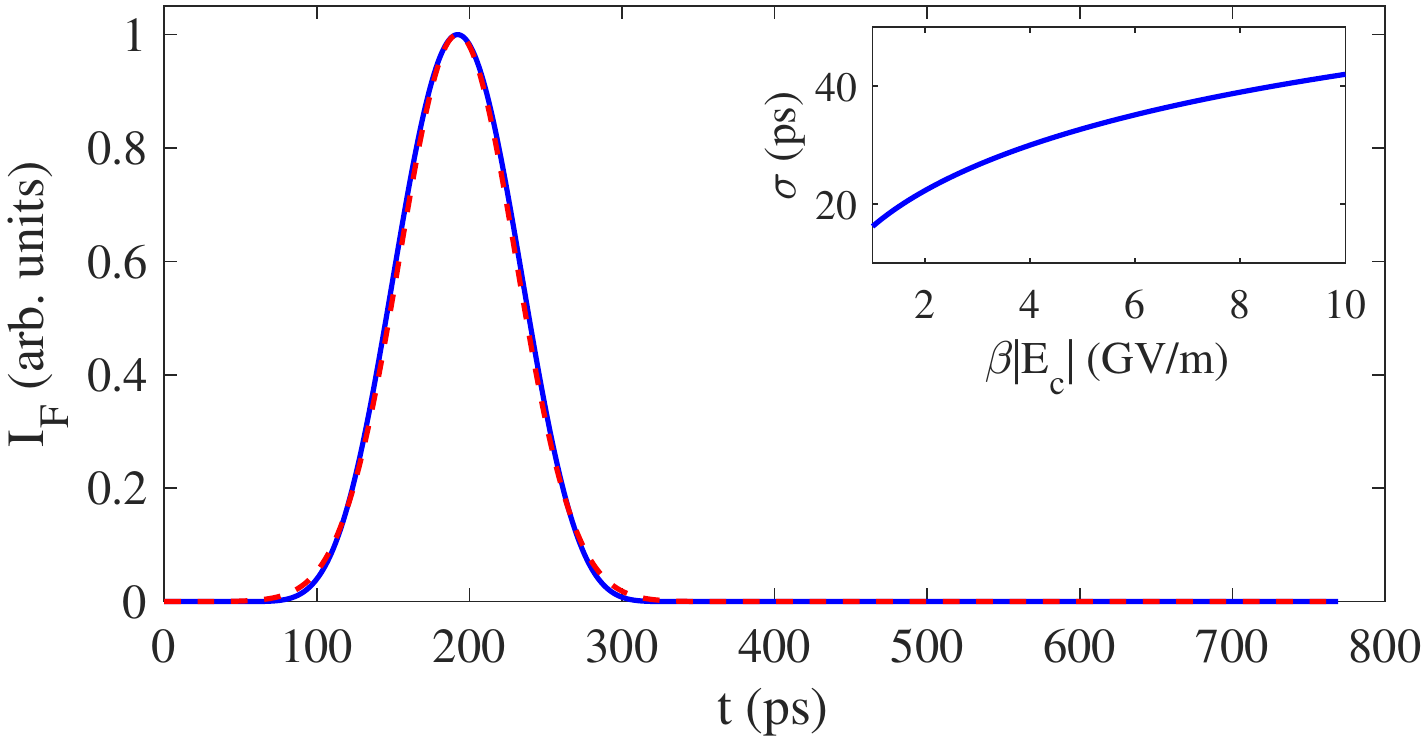}
		\caption{\label{Fig_FE_cycle} Emission profile within one rf cycle based on Eq.~\ref{Eqn_FN_cos} (blue solid line) and its Gaussian distribution approximation (red dashed line). Inset: The standard deviation of the Gaussian distribution as a function the maximum microscopic cathode field. Reproduced from Ref.~\cite{Shao2019}.}
	\end{figure}

	According to the Fowler-Nordheim (FN) equation, the transient field emission current when the cathode field is positive ($\cos (\omega t)>0$) can be expressed as
	\begin{equation}\label{Eqn_FN_cos}
	\begin{aligned}
	I_{F}(t)=&\frac{1.54\times 10^{-6}\times 10^{4.52\phi ^{-0.5}}A_{e}[\beta E_{c}(t)]^{2}}{\phi}\\
	&\times \exp[-\frac{6.53\times 10^{9}\phi ^{1.5}}{\beta E_{c}(t)}]\\
	\end{aligned}
	\end{equation}
	where $\phi$ is the work function. The emission profile can be approximated by a Gaussian distribution whose standard deviation $\sigma$ depends on the maximum microscopic cathode field $\beta |E_{c}|$, as illustrated in Fig.~\ref{Fig_FE_cycle}.
	
	To translate the $Q$-$E$ curves into the Fowler-Nordheim coordinate, first, we determine the pulse length as a function of the local field. In the rf environment, the Fowler-Nordheim coordinates are different than those in the dc environment: one plots $log_{10}(Q/E_c^{2.5})$ as function of $E_c^{-1}$ instead of $log_{10}(Q/E_c^{2})$ as function of $E_c^{-1}$ (like it is in the dc environment). The extra factor of 0.5 is a result of averaging the Fowler-Nordheim current distribution over an rf cycle. The current distribution for Fowler-Nordheim in an rf environment reads\cite{JuwenSLAC1997}
	
	\begin{equation}\label{Eqn_FN_ave}
	\begin{aligned}
	\overline{I_{F}(t)}=&\frac{5.7\times 10^{-12}\times 10^{4.52\phi ^{-0.5}}A_{e}[\beta |E_{c}(t)|]^{2.5}}{\phi ^{1.75}}\\
	&\times \exp[-\frac{6.53\times 10^{9}\phi ^{1.5}}{\beta |E_{c}(t)|}]\\
	\end{aligned}
	\end{equation}
\noindent here the external electric field is modeled as a time variant sinusoidal oscillation which is a result of only considering the longitudinal component as rf guns are TM mode cavity resonators.

\begin{figure}[h!tbp]
		\includegraphics[width=8cm]{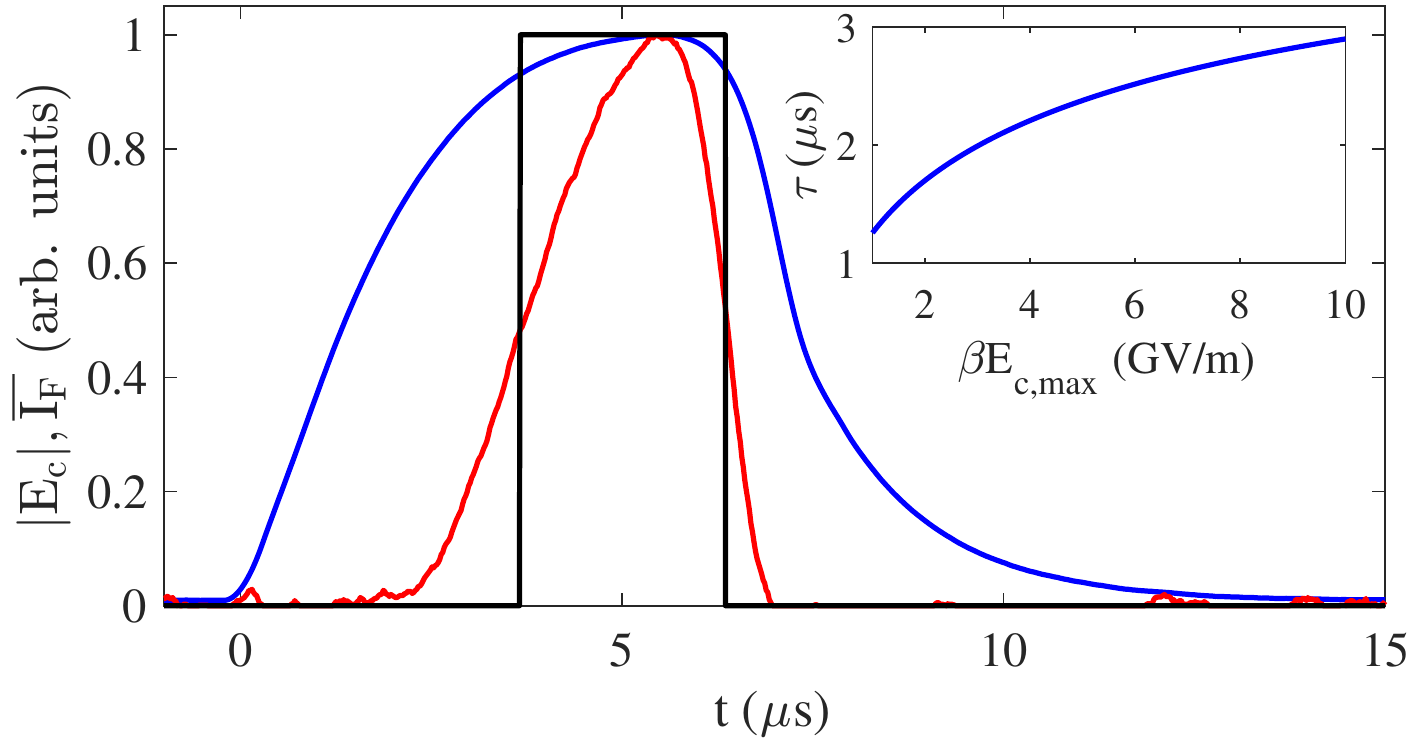}
		\caption{\label{Fig_FE_rf_pulse} Blue: the normalized cathode field amplitude $|E_{c}(t)|$ measured by the rf pickup. Red: the predicted average emission current $\overline{I_{F}(t)}$ by Eq.~\ref{Eqn_FN_ave}. Black: the square pulse approximation of the emission profile. Inset: The width of the square emission profile as a function of $\beta E_{c,\max}$. Reproduced from Ref.~\cite{Shao2019}.}
	\end{figure}

\begin{figure}[htp]
		\includegraphics[width=7cm]{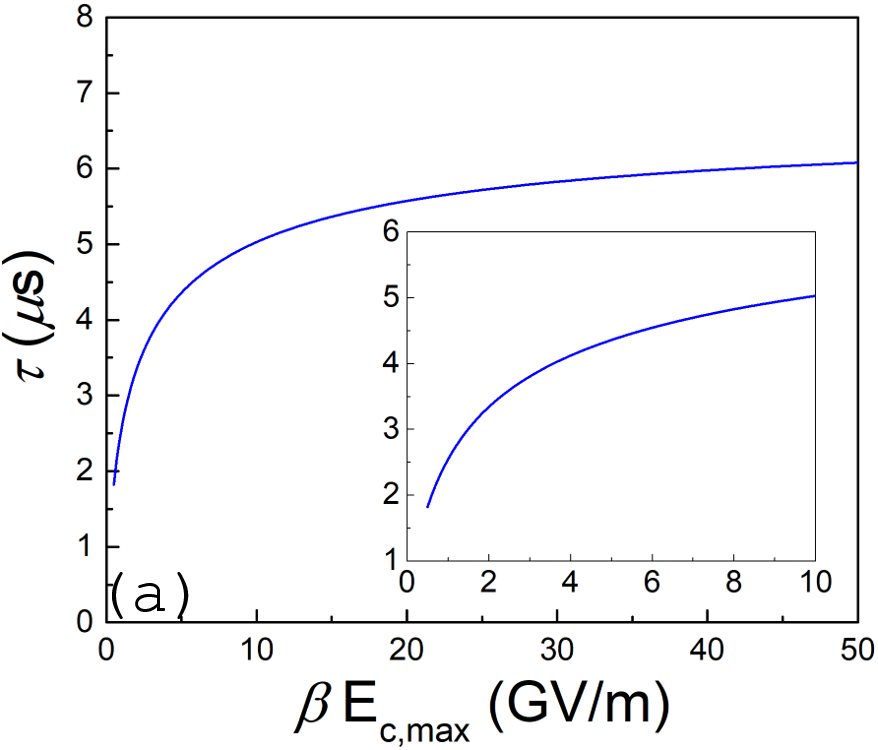}
	\hspace*{0.2cm}\includegraphics[width=7cm]{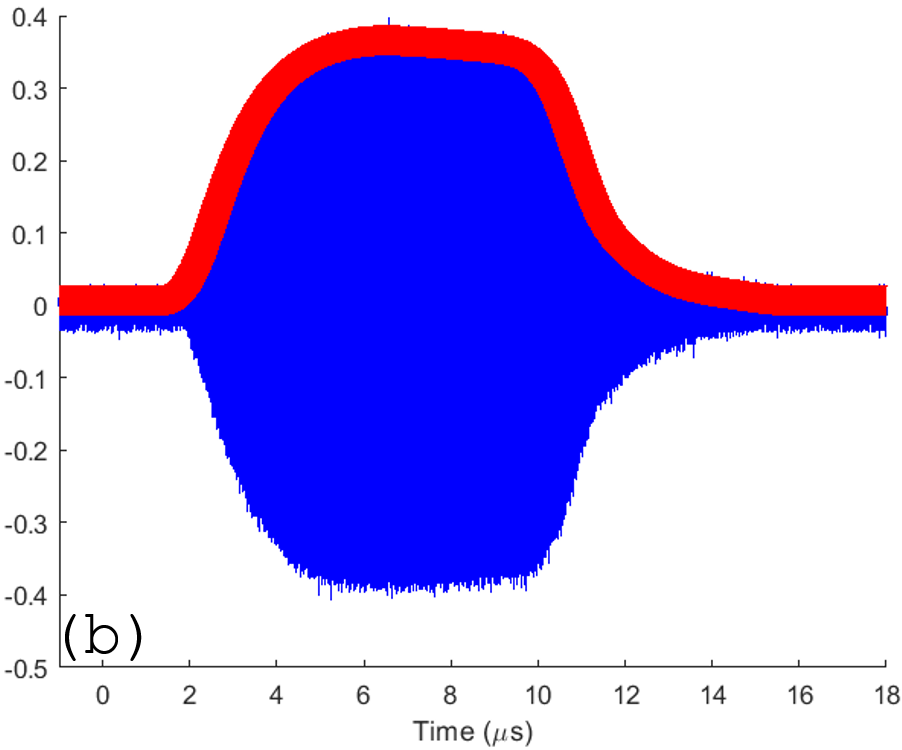}
		\caption{\label{7} (a) The emission pulse pulse length ($\tau$) as a function of the local field plotted in the entire fitting range from 0 to 50 GV/m, inset is a zoom-in view of the pulse length within the physically meaningful local field range, from 0 to 10 GV/m. (b) In red is the rf envelope after frequency filter was applied, in blue is the actual pulse measured using pickup in the gun.}
	\end{figure}

To calculate the emission pulse length, envelope of the drive rf is necessary to be known. Predicted emission profile, as illustrated in Fig.~\ref{Fig_FE_rf_pulse}, is calculated for each $|E_{c}|$, where $\beta$ is assumed to be constant and $\sigma$ of the longitudinal emission profile is adjusted based on $\beta |E_{c}|$. 	It is seen that $\overline{I_{F}(t)}$ is highly sensitive to $|E_{c}|$, and emission pulse can be approximated by using a square emission profile of a length $\tau$ with an average emission current of $\overline{I_{F,\max}}$ calculated by using $|E_{c}|=E_{c,\max}$ in Eq.~\ref{Eqn_FN_ave}, as illustrated in Fig.~\ref{Fig_FE_rf_pulse}. The width of the square pulse is set as $\tau =\int \overline{I_{F}(t)}dt/\overline{I_{F,\max}}$ so as to keep the same charge. It depends on $\beta E_{c,\max}$, as calculated by Eq.~\ref{Eqn_FN_ave} and illustrated in the inset of Fig.~\ref{Fig_FE_rf_pulse}. Note that this calculation routine is done in the wider range of 0-50 GV/m for the local field to allow for convergence, but only the range of 0-10 GV/m is physically meaningful\cite{10giga}, see Fig.~\ref{7}a.

Red line in Fig.~\ref{7}b illustrates rf pulse envelope measurement. Additional smoothing is possible using an appropriate frequency filter function that can be obtained from the same scope. In Matlab, the smoothing is applied using the convolution command called $conv$. Both Figs.~\ref{Fig_FE_rf_pulse} (6 $\mu$s rf pulse) and \ref{7} (8 $\mu$s rf pulse) highlight that the emission pulse length $\tau$ is somewhat shorter than the driving rf pulse length. FEbeam exports the rf envelope and the pulse length as functions of the local field to the figures folder.



\begin{figure}[htp]
		\includegraphics[width=7cm]{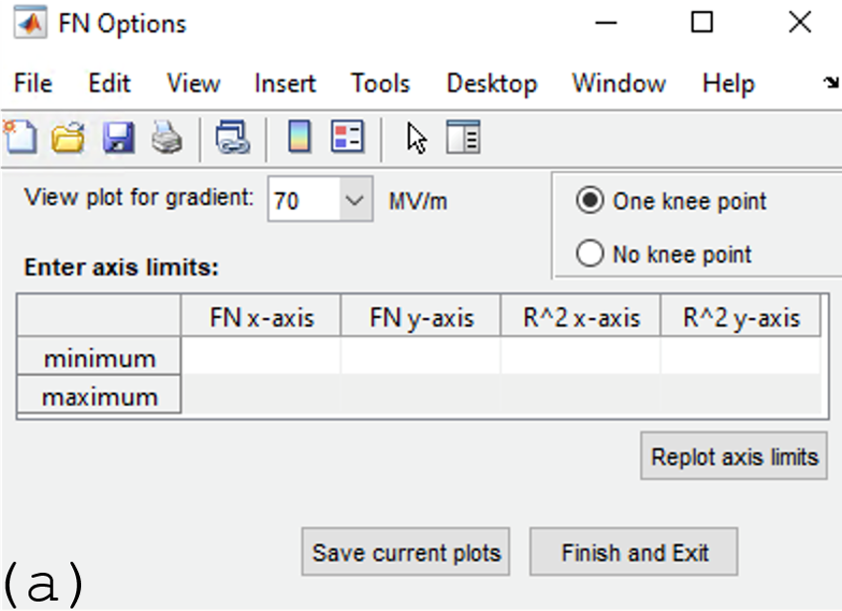}
		\includegraphics[width=7cm]{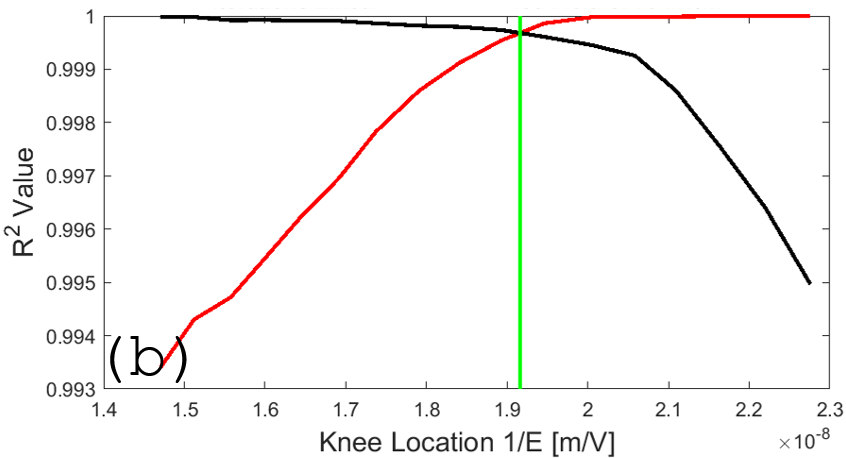}
		\includegraphics[width=7cm]{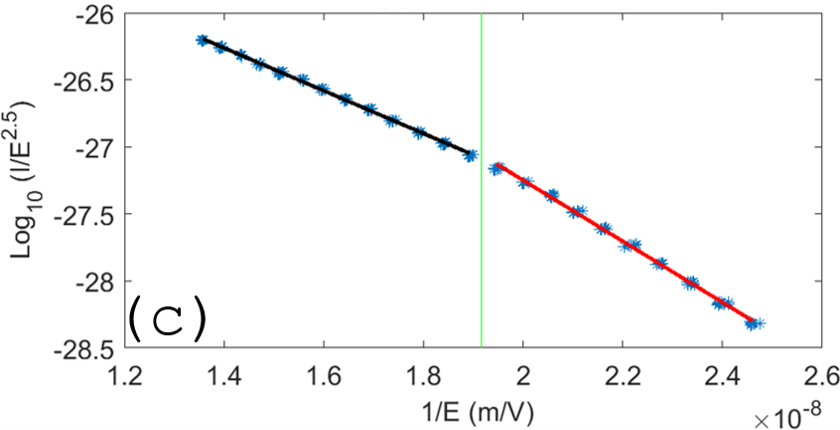}
		\caption{\label{8} (a) Display interface for the Fowler-Nordheim fitting; (b) example of an $R^2$ plot; (c) example of a Fowler-Nordheim plot showing the fitting of the two different regions emphasizing the knee point.}
	\end{figure}

When the pulse length as a function of the local field is found, $Q$-$E$ or $I$-$E$ dependences can be plotted. It is often the case that the field emission characteristic plotted in Fowler-Nordheim coordinates deviates from straight line and multiple slopes can be seen; an example can be seen in Fig.~\ref{8}. There is a number of physical mechanisms at play and this is currently a subject of intense research. Thus, a routine searching for a knee point was developed based on an earlier idea implemented for dc case\cite{Taha}. It is included as an optional data processing tool. The knee point is calculated using the Fowler-Nordheim coordinates of $log_{10}(Q/E_c^{2.5})$ as function of $E_c^{-1}$. Since the pulse length is known to be relatively constant throughout conditioning, it does not change the selection of the knee point location when using the $Q$-$E$ version vs. the $I$-$E$ version of the Fowler-Nordheim coordinates. This only constitutes an additional $y$-offset.

For the ACT, every data point of the $Q$-$E$ measurement consists of 10 pulses (or, shots). Averaged values and an iterative fitting algorithm are used where the Fowler-Nordheim plot is split into two independent line segments shown in Fig.~\ref{8}c. The first line (red line) initially only considers the first three data points and iterates until it fits all but the last three data points. The second line (black line) consists of the remaining points not fit by the first line. The $R^2$ values of each line are recorded for each step in the iteration. The knee point is chosen for the field at which the $R^2$ values of each line intersect, as this is the point where the linear fit is optimized for each line.

If a knee point is found, the Fowler-Nordheim plot is split into two regions. Each region is fit independently of each other to find the $\beta$-factor and emission area $A_e$ for each line segment, i.e. for the high and low field regions. Additionally, the knee point is applied to extract $\beta$-factor and $A_e$ for both the $I$-$E$ and $Q$-$E$ dependences. A comparison between these two fitting methods can be seen using the postprocessing options along with image processing results. Both of these options are independent from the main data processing pipeline and can be done in tandem with the data processing.

If no knee point is found, the entire Fowler-Nordheim plot is fit by a single line. FEbeam displays figures in Fowler-Nordheim coordinates with the knee point indicated. Additional plot showing the $R^2$ value is generated if the knee point was determined. It should be noted that if the algorithm cannot find the knee point due to no diversions from the Fowler-Nordheim law, the user is able to select an option on the FN Options interface (seen in Fig.~\ref{8}a) to fit the $Q$-$E$ data without a knee point. This interface also allows for the user to save the data and figures of the Fowler-Nordheim plot and the $R^2$ plot with the additional ability to change the axis range for better display (Fig.~\ref{8}b and Fig.~\ref{8}c). This can be changed for each conditioning field $E_h$ chosen, and each data set needs to be saved independently by selecting the save button for each $E_h$.

\section{\label{all}Image processing and Postprocessing}

\begin{figure}[htp]
		\includegraphics[width=9cm]{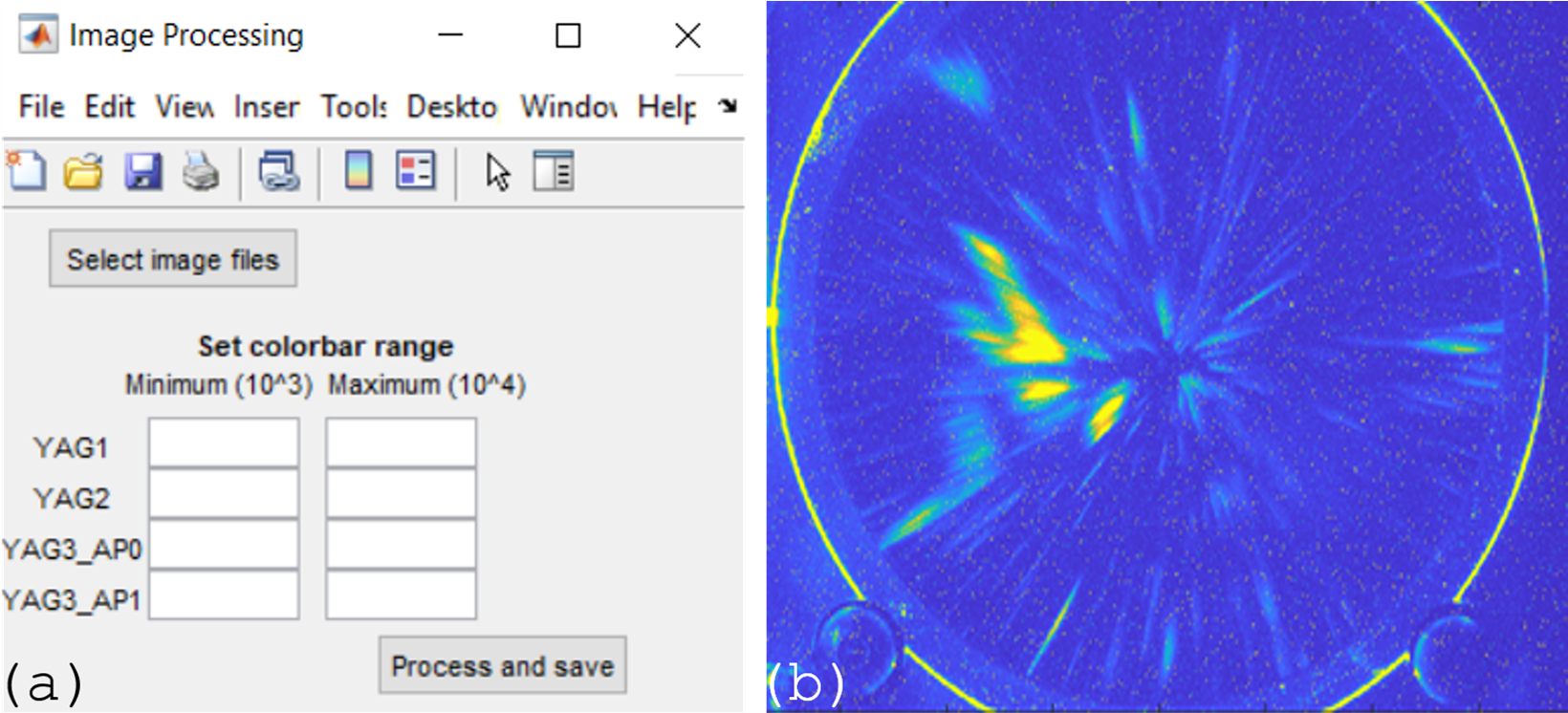}
		\caption{\label{9} (a) Image Processing window showing options to set minimum and maximum intensity threshold for the color bar image. (b) Example image of a YAG3 image with a 1 mm aperture with color bar range 1,000 to 10,000.}
\end{figure}

\begin{figure}[htp]
		\includegraphics[width=9cm]{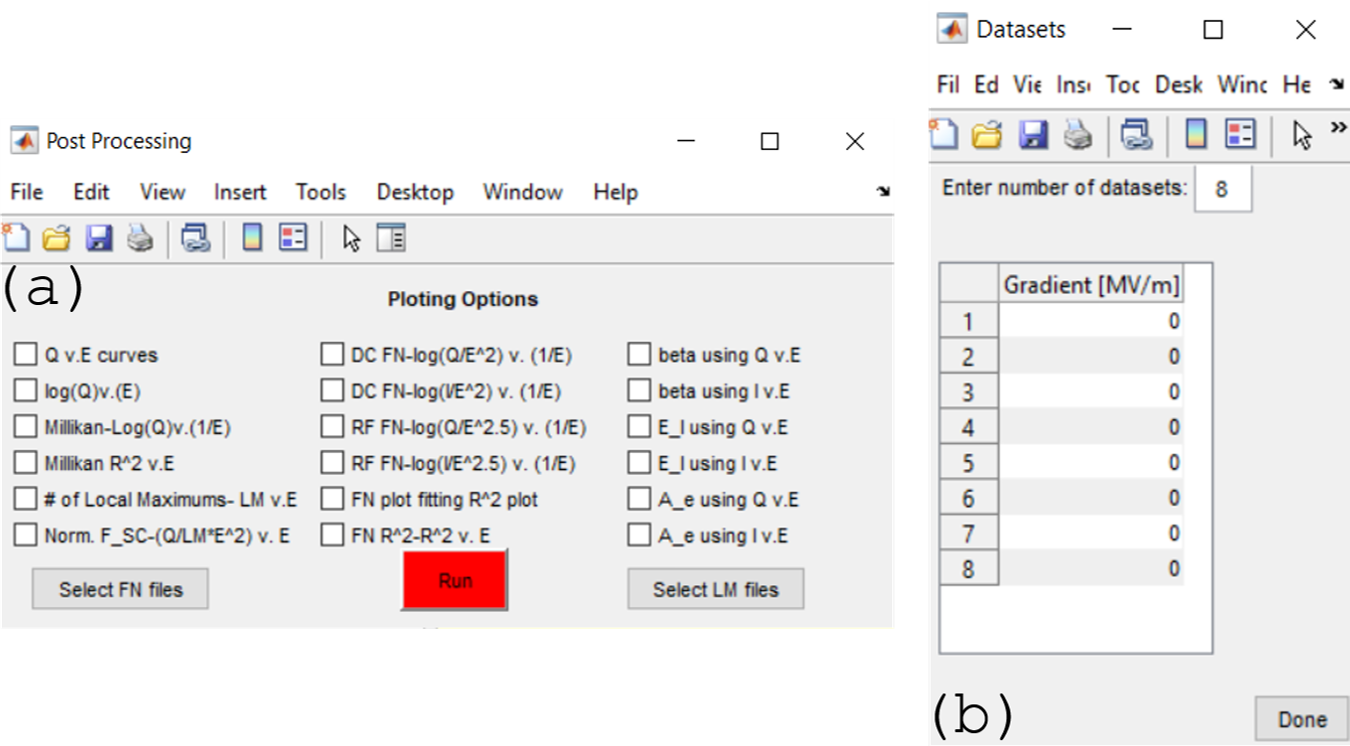}
		\caption{\label{10} (a) Post Processing window showing all available postprocessing options; (b) pop-up window that allows the user to select which gradients they would like to postprocess.}
\end{figure}

The image processing option in FEbeam is self-contained from the rest of the data processing pipeline seen in Fig.~\ref{1}. If the image processing option is selected first, the operator imports the raw image files in the format of .DAT. The user selects the minimum and maximum pixel values which are then applied to all of the images to be processed as the color bar. As this image processing option (see Fig.~\ref{9}) was specified for the ACT, there are options for the three YAG screens on the ACT. The YAG3 images can be processed with and without the 1 mm aperture (located at the position of YAG2). The exported images are in the standard RGB color map as found in Matlab, where the limits of the color bar are set by the inputs (exampled in Fig.~\ref{9}a), and an pseudo-color example post-processed image is shown in Fig.~\ref{9}b. These results can then be coupled to the variety of postprocessing options as illustrated in Fig.~\ref{10}a and Fig.~\ref{10}b, allowing the user to pick which conditioning fields are postprocessed.

Beyond the traditional Fowler-Nordheim parameters and Fowler-Nordheim plot, the fitting can also be compared in both the $I$-$E$ and $Q$-$E$ versions. Additionally, the dc Fowler-Nordheim plot $log(I/E^{2})$ vs. $1/E$ is available. There are options for the Millikan plot $log(I)$ or $log(Q)$ vs. $1/E$ and its $R^2$ values which gives the user better insight if significant space charge effect is present\cite{SC}. The normalized space charge force can be plotted after importing the number of emitters obtained from the optional image reconginition module FEpic. FEpic calculates the number of field emission centers for each image in a separate independent image processing pipeline\cite{FEpic}.

\section{\label{conc}Conclusions and Outlook}
As field emission sources are notoriously difficult to characterize, FEbeam provides an intuitive field emission data processing pipeline with the additional benefit of having image and post processing capabilities. FEbeam is also of a highly modular design with only one subsection being specific to the ACT, which it was originally designed for. Supplementary modules can be added for additional functionality. 
New work is underway aiming to join FEbeam, FEpic (expanded to unsupervised machine learning algorithm), and a breakdown analyses module FEbreak under a single umbrella called FEmaster.
	
		\begin{acknowledgments}
		This work was supported by the US Department of Energy, Office of Science, High Energy Physics under Cooperative Agreement award No. DE-SC0018362. This material is also based upon work supported by the U.S. Department of Energy, Office of Science, Office of High Energy Physics under Award No. DE-SC0020429. Partial support was provided under the Global Impact Initiative (College of Engineering, Michigan State University). The work at AWA is funded through the U.S. Department of Energy Office of Science under Contract No. DE-AC02-06CH11357.
	\end{acknowledgments}

	\section*{References}
	\bibliography{FEbeam}

\section*{Appendix A}
\noindent Program versions:
\begin{itemize}
  \item Matlab version R2019b
  \item Python version 2.7 $\times$86-64 MSI
\end{itemize}

\noindent Follow documentation \href{https://www.mathworks.com/help/matlab/matlab_external/install-supported-python-implementation.html}{HERE} to configure Matlab to interface with Python (required for function).

In order for FEbeam to function and save files properly, data files must be stored in the exact folder format shown below. Bolded folder names indicate that exact name must be used in that location to ensure that FEbeam can locate where to store files it creates. The folder for each day must contain the folders as shown below

\

\

\begin{itemize}
   \item Experiment
   \begin{itemize}
     \item Day
     
     \begin{itemize}
       \item \textbf{raw}
       \item \textbf{raw matlab}
       \item \textbf{settings}
       \item \textbf{EC\_info}
       \item \textbf{image}
       \begin{itemize}
       \item DAT
       \item \textbf{png}
       \end{itemize}
     \end{itemize}
     \item \textbf{figures}
     \item \textbf{all IV curves}
     \item \textbf{FN data}
   \end{itemize}
 \end{itemize}

\begin{center}
\newcolumntype{A}{ >{\centering\arraybackslash} m{3cm} }
\newcolumntype{B}{ >{\centering\arraybackslash} m{2.5cm} }
\begin{tabular}{ A | B | B} 
\hline \hline
\vspace{3pt}
\textbf{Data Processing Input Files} & \multirow{-2}{*}{\textbf{File Format}} & \multirow{-2}{*}{\textbf{File Location}} \\

\hline
\vspace{3pt}
Raw data* & .CSV & raw \\

\hline
\vspace{3pt}
Parameters (low charge case only) & \multirow{-2}{*}{.CSV} & \multirow{-2}{*}{raw} \\

\hline
\vspace{3pt}
Time constant & .MAT & settings \\

\hline
\vspace{3pt}
rf filter & .MAT & experiment \\

\hline
\vspace{3pt}
rf envelope & .DAT & experiment \\

\hline \hline
\end{tabular}
\end{center}

\

\noindent *Raw data filenames must contain the timestamp that corresponds to the gradient value entered in the window shown in Fig.~\ref{3}b.

\begin{center}
\newcolumntype{A}{ >{\centering\arraybackslash} m{3cm} }
\newcolumntype{B}{ >{\centering\arraybackslash} m{2.5cm} }
\begin{tabular}{ A | B | B } 
\hline \hline
\vspace{3pt}
\textbf{Image Processing Input Files} & \multirow{-3}{*}{\textbf{File Format}} & \multirow{-3}{*}{\textbf{File Location}} \\
\hline
\vspace{3pt}
Image files & .DAT & DAT \\
\hline \hline
\end{tabular}
\end{center}

\begin{center}
\newcolumntype{A}{ >{\centering\arraybackslash} m{3cm} }
\newcolumntype{B}{ >{\centering\arraybackslash} m{2.5cm} }
\begin{tabular}{ A | B | B } 
\hline \hline
\vspace{3pt}
\textbf{Postprocessing Input Files} & \multirow{-2}{*}{\textbf{File Format}} & \multirow{-2}{*}{\textbf{File Location}} \\
\hline
\vspace{3pt}
FN files** & .MAT & FN data \\
\hline
\vspace{3pt}
Local maxima files & .CSV & $\sim$ \\
\hline \hline
\end{tabular}
\end{center}

\

\noindent **File grouping and numbering is done by the maximal gradient $E_h$ achieved in the given completed experimental cycle. If duplicate gradient values are being postprocessed (say, when an experiment gets repeated for the sake of reproducibility), the experiment number or experimental phase must be denoted in the filename by using a decimal. For example, if two 70 MV/m files are being used, one filename must contain 70.1 and the other must contain 70.2, denoting phase/experiment 1 and phase/experiment 2 respectively. FEbeam allows for a maximum of 9 phases at once for a single gradient (phases 1-9). In the window shown in Fig.~\ref{10}b, if phases are used, the gradient values must be entered as 70.1 and 70.2, per this example case. Plots are titled accordingly, such as containing ``70 MV/m Phase 1'' and ``70 MV/m Phase 2''. If duplicate gradients are used in postprocessing, the phase must be denoted in this manner for all duplicate values to ensure proper functionality. 	
\end{document}